\documentclass[conference]{IEEEtran}
\IEEEoverridecommandlockouts
\usepackage{cite}
\usepackage{babel}
\usepackage{amsmath,amssymb,amsfonts}
\usepackage{algorithmic}
\usepackage{graphicx}
\usepackage{textcomp}
\usepackage{xcolor}

\newcommand{\ing}[1]{\mathsf{#1}}
\newcommand{\Rn}[1]{\ifthenelse{\equal{#1}{}}{\mathbb{R}}{\mathbb{R}^{\ing{#1}}}}
\newcommand{\Cn}[1]{\mathbb{C}^{\ing{#1}}}
\newcommand{\Rset}[2]{\ifthenelse{\equal{#2}{1}}{\in \Rn{\ing{#1}}}{\in \Rn{\ing{#1} \times \ing{#2}}}}
\newcommand{\Cset}[2]{\ifthenelse{\equal{#2}{1}}{\in \Cn{\ing{#1}}}{\in \Cn{\ing{#1} \times \ing{#2}}}}
\newcommand{\vect}[1]{\boldsymbol{\mathbf{\MakeLowercase{#1}}}}
\newcommand{\mtrx}[1]{\boldsymbol{\mathbf{#1}}}

\newcommand{\norm}[2]{\|#1\|_{#2}}

\newcommand{\transp}[1]{#1^{\mathsf{T}}}

\newcommand{\ind}[2]{\ifthenelse{\equal{#2}{}}{\chi^{\text{DP}}_{#1}}{\chi^{\text{DP}}_{#1}\left( #2 \right)}}
\newcommand{\diag}[1]{{\mathtt{diag}}\left(#1\right)}
\newcommand{\iter}[2]{#1(\ing{#2})}
\DeclareMathOperator{\vtriu}{vtriu}
\DeclareMathOperator*{\argmin}{argmin}

\def\BibTeX{{\rm B\kern-.05em{\sc i\kern-.025em b}\kern-.08em
    T\kern-.1667em\lower.7ex\hbox{E}\kern-.125emX}}

\begin{document}

\title{Echo-enabled Direction-of-Arrival and range estimation of a mobile source in Ambisonic domain}

\author{\IEEEauthorblockN{J\'er\^ome Daniel}
\IEEEauthorblockA{\textit{Orange Labs} \\
\textit{Cesson-S\'evign\'e, France}\\
jerome.daniel@orange.com}
\and
\IEEEauthorblockN{Sr\dj{}an Kiti\'c}
\IEEEauthorblockA{\textit{Orange Labs} \\
\textit{Cesson-S\'evign\'e, France}\\
srdan.kitic@orange.com}\thanks{The two authors have equally contributed to the present article.}}

\maketitle

\begin{abstract}
Range estimation of a far field sound source in a reverberant environment is known to be a notoriously difficult problem, hence most localization methods are only capable of estimating the source’s Direction-of-Arrival (DoA). In an earlier work, we have demonstrated that, under certain restrictive acoustic conditions and given the orientation of a reflecting surface, one can exploit the dominant acoustic reflection to evaluate the DoA \emph{and} the distance to a static sound source in Ambisonic domain. In this article, we leverage the recently presented Generalized Time-domain Velocity Vector (GTVV) representation to estimate these quantities for a moving sound source without an a priori knowledge of reflectors’ orientations. We show that the trajectories of a moving source and the corresponding reflections are spatially and temporally related, which can be used to infer the absolute delay of the propagating source signal and, therefore, approximate the microphone-to-source distance. Experiments on real sound data confirm the validity of the proposed approach.
\end{abstract}


\section{Introduction}

Being an important preprocessing task in many applications (including, \emph{e.g.} multichannel speech separation \cite{gannot2017consolidated,bosca2021dilated}, 
robot navigation \cite{rascon2017localization}, %,an2021diffraction,valin2007robust},
spatial audio coding \cite{pulkki2007spatial} %,daniel2003spatial}
and reproduction \cite{kowalczyk2015parametric}), sound source localization has been a longstanding, yet still very active area of research in audio signal processing. Due to numerous acoustic phenomena that occur in real environments, localization remains to be a challenging problem. The presence of noise and reverberation can have detrimental influence on many traditional signal processing methods, by degrading their localization accuracy to various degrees \cite{blandin2012multi}. Reverberation, in particular, is usually considered a significant nuisance, thus a lot of research effort has been devoted to combating its effects, either by explicitly avoiding/suppressing reverberate portions of a signal \cite{madmoni2018direction}, or by using robust trained models, such as deep neural networks \cite{grumiaux2021survey}. Conversely, a number of works in the last decade has exercised the possibility of exploiting the multipath propagation of acoustic waves for the purpose of, \emph{e.g.} signal enhancement %\cite{dokmanic2015raking,
\cite{scheibler2018separake}, source localization behind soundproof obstacles \cite{an2021diffraction}, %,kitic2014hearing}, 
room geometry estimation \cite{dokmanic2013acoustic} %,mabande2013room}, 
and simultaneous localization and mapping (SLAM) \cite{krekovic2016echoslam}. 

In \cite{daniel2020time}, we have shown that one could infer the 3D position of a static sound source with respect to the Ambisonic \cite{rafaely2018fundamentals,jarrett2017theory} microphone array (\emph{i.e.}, its DoA and range), given an estimate of the direction and relative delay of the acoustic reflection coming from a horizontal surface. Moreover, it has been shown that such information can be efficiently extracted from the \emph{Time Domain Velocity Vector} (\emph{TDVV}, \emph{a.k.a.} Relative Impulse Response in spherical harmonics domain \cite{jarrett2017theory,gannot2017consolidated}), albeit only under very restrictive acoustic conditions. Recently, we have introduced an enriched version of this spatial representation, the so-called \emph{Generalized Time-domain Velocity Vector (GTVV)} \cite{kitic2021generalized}, that improves upon TDVV in two ways. First, it incorporates Higher Order Ambisonic (HOA) channels, which increase the spatial resolution of the wavefronts that can be identified from GTVV, and second, its reference component is a linear combination of \emph{all} available HOA channels (whilst in TDVV, it was restricted to the omnidirectional channel only). A convenient choice of this reference enables GTVV to preserve useful theoretical properties even in adverse acoustic conditions.

In this work our goal is to exploit the GTVV imprint in order to estimate the 3D position of a moving sound source, without strong assumptions concerning the orientations of reflecting surfaces. Since the source is mobile, the inference should be done at the frame rate, whereas, in \cite{daniel2020time}, a single range estimate has been issued for the entire Ambisonic recording. We show that the moving source provides a ``spatial diversity'' that will be leveraged to cope with the unknown geometry of the acoustic environment. In principle, the discussed ideas could be transposed to the setting where the source is fixed while the microphone is mobile, due to the problem symmetry.

The article is organized as follows. In Section~\ref{sec:GTVV}, we briefly recall the GTVV representation, its estimation, properties and the inference of individual wavefronts' parameters. Next, in Section~\ref{sec:range}, we discuss the distance estimation using such extracted information. We demonstrate the validity of the proposed method on real sound data in Section~\ref{sec:experiments}. The conclusion is given in Section~\ref{sec:conclusion}.

\section{The GTVV representation}\label{sec:GTVV}

Let $\vect{b}(f) \Cset{(L+1)^2}{1}$ denote the vector whose entries are concatenated Ambisonic channels of orders $\ing{l} \in [0,\ing{L}]$, and degrees $\ing{m} \in [-\ing{l},\ing{l}]$, at frequency $f$ \cite{rafaely2018fundamentals}. Assuming the presence of a single, dominant source signal $S(f)$, one can write

\begin{equation}
    \vect{b}(f) = S(f) \sum\limits_{\ing{n}=0}^{\ing{N}} a_{\ing{n}}(f) \vect{y}_{\ing{n}} + \vect{n}(f),
\end{equation}
where the summation term models the early echoes, while $\vect{n}(f)$ captures late reverberation and additive noise. The vector $\vect{y}_{\ing{n}}$ contains real-valued spherical harmonic coefficients  \cite{jarrett2017theory} $\{ Y_{\ing{lm}}(\Omega_{\ing{n}}) \}_{\ing{l},\ing{m}}$, corresponding to a wavefront coming from the direction  ${\Omega_{\ing{n}}:= (\theta_{\ing{n}}, \varphi_{\ing{n}})}$, with $\theta_{\ing{n}}$ and $\varphi_{\ing{n}}$ being the azimuth and elevation, respectively. The multiplicative factors ${a_{\ing{n}}(f) = h_{\ing{n}}(f) e^{-j2 \pi f \bar{\tau}_{\ing{n}}}}$ capture the magnitude and phase shift of individual wavefronts (the delay $\bar{\tau}_{\ing{n}}$ is the Time-of-Arrival (ToA) of the $\ing{n}$\textsuperscript{th} wavefront).

In the ``noiseless'' case, the Generalized \emph{Frequency} domain Velocity Vector (GFVV) has been defined \cite{kitic2021generalized} as
\begin{equation}\label{eqFDVVinst}
    \vect{v}(f) = \frac{\vect{b}(f)}{\transp{\vect{w}(f)}\vect{b}(f) } = \frac{ \sum\limits_{\ing{n}=0}^{\ing{N}} a_n(f) \vect{y}_{\ing{n}}}{ \sum\limits_{\ing{n}=0}^{\ing{N}} a_{\ing{n}}(f) 
    \beta_{\ing{n}}(f)} = \frac{\vect{y}_0 + \sum\limits_{\ing{n}=1}^{\ing{N}} \gamma_{\ing{n}}(f) \vect{y}_{\ing{n}} }{1 + \sum\limits_{\ing{n}=1}^{\ing{N}} \gamma_{\ing{n}}(f)\beta_{\ing{n}}(f) },
\end{equation}
where $\vect{w}(f)$ is a complex weight vector (in this work, a maximum-directivity beamformer \cite{jarrett2017theory} $\vect{w}(f) := \vect{w}$ steered roughly towards DoA), and $\beta_{\ing{n}}(f)=\beta_{\ing{n}} =\transp{\vect{w}}\vect{y}_{\ing{n}} \leq 1$. The rightmost expression is obtained by factoring out $a_0(f)$, which gives $\gamma_{\ing{n}} = g_{\ing{n}}(f) e^{-j2\pi f (\bar{\tau}_{\ing{n}} - \bar{\tau}_0)}$, with $g_{\ing{n}}(f) = h_{\ing{n}}(f)/h_{\ing{0}}(f)$ being the relative gain of the $\ing{n}$\textsuperscript{th} wavefront with respect to the wavefront propagating from the DoA. Provided that $S(f)$ is non-negligible at the given frequency $f$, GFVV depends solely on the spatio-temporal distribution of contributing wavefronts (and not on the source signal content). 

In practice, the recorded signals $\vect{b}(f)$ are never noise-free, thus the GFVV cannot be computed by directly applying the expression \eqref{eqFDVVinst}. Instead, one should use some noise-robust estimator, such as the adaptation of the relative transfer function estimator %\cite{gannot2001signal,
\cite{jarrett2017theory,gannot2017consolidated}, proposed in \cite{kitic2021generalized}. In the latter case, the multichannel Ambisonic signal is represented in time-frequency (namely, the Short-Time Fourier Transform (STFT)) domain $\vect{b}(\ing{k},f) = \transp{\left[ \begin{smallmatrix} b_{00}(\ing{k},f) & b_{1,-1}(\ing{k},f) & \hdots & b_{\ing{LL}}(\ing{k},f) \end{smallmatrix} \right]}$, where $\ing{k}$ denotes the frame index. Assuming that the source does not change its position significantly within $\ing{T}$ frames, each $(\ing{l},\ing{m})$ entry $v_{\ing{lm}}(f)$ of the vector $\vect{v}(f)$, along with the noise cross-spectrum estimate $\sigma_{\ing{lm}}(f)$, is the solution of the following overdetermined system of equations:
\begin{equation}\label{eqRobust}
  \vect{\phi}_{\ing{lm}}(f) = \left[ \begin{matrix} \mtrx{\Phi}_{\ing{lm}}(f)\vect{w} & \vect{1} \end{matrix} \right]\left[ \begin{matrix} v_{\ing{lm}}(f) \\ \sigma_{\ing{lm}}(f) \end{matrix} \right].
\end{equation}
Above, the vector $\vect{\phi}_{\ing{lm}}(f)$ contains (weighted) power spectrum estimates of the $\ing{T}$ frames prior to the current frame $\ing{k}$:
\begin{equation*}
\vect{\phi}_{\ing{lm}}(f) = \left[ \begin{matrix} \tilde{b}_{\ing{lm}}(\ing{k},f)\tilde{b}_{\ing{lm}}(\ing{k},f)^* \\ 
\tilde{b}_{\ing{lm}}(\ing{k-1},f)\tilde{b}_{\ing{lm}}(\ing{k-1},f)^* \\
\vdots \\
\tilde{b}_{\ing{lm}}(\ing{k-T+1},f)\tilde{b}_{\ing{lm}}(\ing{k-T+1},f)^* \end{matrix} \right],
\end{equation*}
where $\tilde{b}_{\ing{lm}}(\ing{k},f) := \omega(\ing{k},f) b_{\ing{lm}}(\ing{k},f)$, with $\omega(\ing{k},f)$ being the time-frequency weights.

Accordingly, $\mtrx{\Phi}_{\ing{lm}}(f)$ is the matrix of weighted cross-spectrum values of all Ambisonic channels $(\ing{l',m'})$ with respect to the considered channel $\ing{l,m}$:
\begin{equation*}
    \mtrx{\Phi}_{\ing{lm}} = \left[ \begin{matrix} \tilde{b}_{\ing{00}}(1) \tilde{b}_{\ing{lm}}(1)^* & \hdots & \tilde{b}_{\ing{LL}}(1) \tilde{b}_{\ing{lm}}(1)^* \\
\tilde{b}_{\ing{00}}(2) \tilde{b}_{\ing{lm}}(2)^* & \hdots & \tilde{b}_{\ing{LL}}(2)b_{\ing{lm}}2)^* \\
\vdots & \vdots & \vdots \\
\tilde{b}_{\ing{00}}(\ing{T}) \tilde{b}_{\ing{lm}}(\ing{T})^* & \hdots & \tilde{b}_{\ing{LL}}(\ing{T}) \tilde{b}_{\ing{lm}}(\ing{T})^*  \end{matrix} \right],
\end{equation*}
where we have dropped the variable $f$, and used frame indexing $\ing{k}+1 - [1, \; 2, \; \hdots \; \ing{T}]$, for ease of notation. 

The weights $\omega(\ing{k},f)$ inject prior knowledge about the time-frequency distribution of the signal. For instance, these can represent the entries of the mask generated by a source separation algorithm, which may enable processing in the multiple source scenario. In this article, we investigate only the single source case, and apply the weights emphasizing successive time-frequency bins that contain similar information. The intuition is that such regions are akin to sound attacks, which are less likely to be corrupted by the reverberation ``tail'' from the preceding frames. Under the common hypothesis that $b_{\ing{lm}}(\ing{k},f)$ are zero-mean, we compute the weights as Pearson's correlation coefficients (\emph{i.e.} the normalized cross-correlation) between $\vect{b}(\ing{k},f)$ and $\vect{b}(\ing{k}+1,f)$:
\begin{equation*}
    \omega(\ing{k},f) = \frac{\langle \vect{b}(\ing{k},f), \vect{b}(\ing{k}+1,f) \rangle}{\norm{\vect{b}(\ing{k},f)}{} \norm{\vect{b}(\ing{k}+1,f)}{} },
\end{equation*}
where $\langle \cdot, \cdot \rangle$ denotes the scalar product (the normalization reduces the influence of energy variations in the source signal).

Assuming the frequency-independent gains $g_{\ing{n}}(f)=g_{\ing{n}}$, and ${\sum_{\ing{n}=1}^{\ing{N}} |g_{\ing{n}} \beta_{\ing{n}}| < 1}$ (the reflections are strongly attenuated by the beamformer), it can be shown \cite{kitic2021generalized} that the time-domain version of \eqref{eqFDVVinst}, \emph{i.e.}, the GTVV, admits the following expression:
\begin{multline}\label{eqGTVV}
    \vect{v}(t) = \delta(t) \vect{y}_0 +  \\
    \sum\limits_{\ing{p}\geq 1}\sum\limits_{\ing{n}=1}^{\ing{N}} (-g_{\ing{n}}\beta_{\ing{n}})^{\ing{p}} \left( \vect{y}_0 - \frac{1}{\beta_{\ing{n}}} \vect{y}_{\ing{n}} \right)\delta(t - \ing{p}(\bar{\tau}_{\ing{n}} - \bar{\tau}_0)) + \tilde{\eta}(t).
\end{multline}
Here we focus on the terms that depend only on individual wavefronts, encapsulating in $\tilde{\eta}(t)$ the so-called ``cross-terms'', that arise due to wavefronts' mutual interference.%\footnote{An interested reader may consult \cite{GTVV} for the in-depth derivation.}

The GTVV representation allows to straightforwardly determine the DoA component $\vect{y}_0$, by simply evaluating the sequence \eqref{eqGTVV} at $\vect{v}(t=0)$. Moreover, for the strong reflections outside the main lobe of the beamformer ($\beta_{\ing{n}} \ll 1$), we have ${\vect{v}(\bar{\tau}_{\ing{n}} - \bar{\tau}_0) \approx g_{\ing{n}} \vect{y}_{\ing{n}}}$. This allows us to identify few important reflections by selective peak-picking, with the peak index corresponding to the Time Difference-of-Arrival (TDoA) ${\tau_{\ing{n}}=\bar{\tau}_{\ing{n}} - \bar{\tau}_0}$ of a given reflection. The main advantage of such rudimentary procedure is computational efficiency, however, one may opt to more sophisticated inference techniques, \emph{e.g.}, based on Simultaneous Orthogonal Matching Pursuit \cite{tropp2006algorithms}. 

%However, retrieving the actual direction $\hat{\Omega}_{\ing{n}}$ from an observed vector $\hat{\vect{y}}_{\ing{n}}$ is not trivial, due to microphone array imperfections, spatial aliasing, encoding format etc. A simple, though suboptimal way is to compare $\hat{\vect{y}}_{\ing{n}}$ with the dictionary of ``ideal'' steering vectors $\{ \vect{y}_{\ing{d}} \}_{\ing{d}}$, defined on a discrete grid, and choosing the direction $\hat{\Omega}_{\ing{n}}=\Omega_{\ing{d}}$ corresponding to the atom $\vect{y}_{\ing{d}}$ that is maximally aligned with $\hat{\vect{y}}_{\ing{n}}$.

\section{Range estimation}\label{sec:range}

The following discussion is based on the well-known image source model of sound wave propagation \cite{allen1979image}, and, thus, an implicit assumption that all involved reflections are specular. This enables us to use simple geometric arguments to arrive at the range (distance) estimate $\iter{d}{k}$, at the frame indexed by $\ing{k}$  (given the speed of sound $c$, this is equivalent to estimating the per-frame ToA $\iter{\bar{\tau}}{k}$, since we have ${\iter{d}{k} = c\iter{\bar{\tau}}{k}}$). Indeed, the image sources are obtained from the original source positions by applying certain rigid transformations: reflections, translations or rotations (depending on the order and the arrangement of the reflecting surfaces). Regardless of their type, the key concept we are about to use is that the rigid transformations preserve Euclidean distances.

%(regardless of their order, \emph{i.e.}, it could be a wavefront that has ``bounced off'' multiple surfaces)

\begin{figure}
    \centering
    \includegraphics[width=\columnwidth]{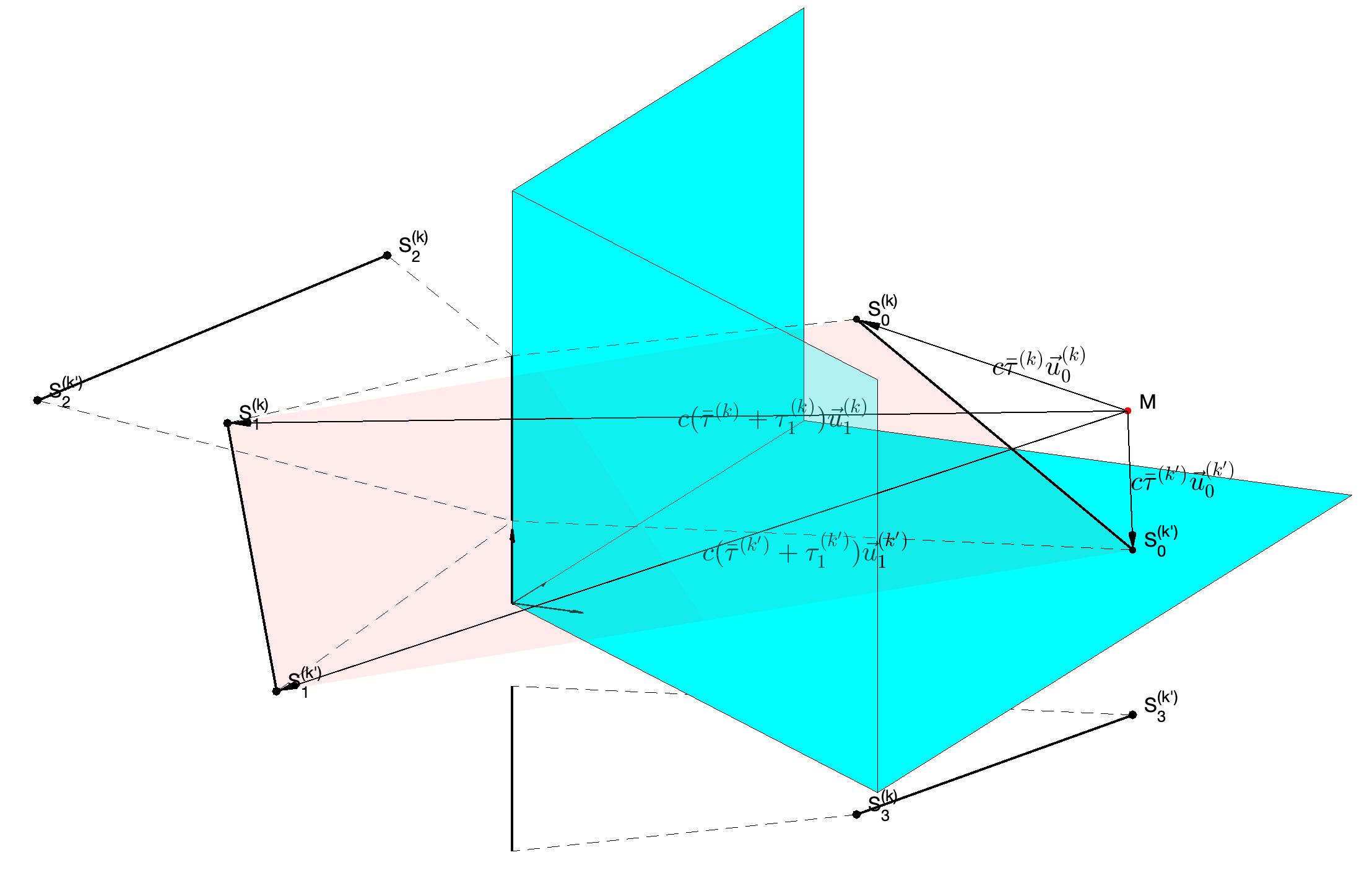}
    \caption{An illustration of the preservation of distance, and of the magnitude of the projection onto the z-axis (point $\text{M}$ is the microphone position).}
    \label{figDPHV}
\end{figure}

We assume hereafter that the method discussed in Section~\ref{sec:GTVV} provides us with the source DoA in the form of a unitary vector $\iter{\vec{u}_0}{k}$, and a collection of pairs $\{(\iter{\vec{u}_{\ing{n}}}{k}, \; \iter{\tau_{\ing{n}}}{k} )\}_{\ing{n}}$ corresponding to the detected reflections and their associated TDoAs, at some frame $\ing{k}$. From Fig.~\ref{figDPHV}, we have that the position of the source relative to the microphone array is given by the vector
\begin{equation}\label{eq_r0}
\iter{\vec{r}_0}{k} = c\iter{\bar{\tau}}{k} \iter{\vec{u}_0}{k}.    
\end{equation}
Likewise, the position of the image source $\ing{n}=1$ is given by
\begin{equation}\label{eq_rn}
    \iter{\vec{r}_{\ing{n}}}{k} = c(\iter{\bar{\tau}}{k} + \iter{\tau_{\ing{n}}}{k}) \iter{\vec{u}_{\ing{n}}}{k}.
\end{equation}
The equivalent expressions for the position vectors $\iter{\vec{r}_0}{k'}$ and $\iter{\vec{r}_{\ing{1}}}{k'}$ would be obtained for the same reflection detected in another frame $\ing{k'}$.

Due to the \emph{distance preservation (DP)} property, we have
\begin{equation}\label{eqDistPreserv}
    \norm{\iter{\vec{r}_0}{k} - \iter{\vec{r}_0}{k'}}{}^2 = \norm{\iter{\vec{r}_{\ing{n}}}{k} - \iter{\vec{r}_{\ing{n}}}{k'}}{}^2.
\end{equation}
Developing this expression using \eqref{eq_r0} and \eqref{eq_rn} yields
\begin{multline}\label{eqDP}
    \iter{\alpha^{\text{DP}}_{\ing{n}}}{k,k'}\iter{\bar{\tau}}{k} + \iter{\alpha^{\text{DP}}_{\ing{n}}}{k',k}\iter{\bar{\tau}}{k'} + \iter{\chi^{\text{DP}}_{\ing{n}}}{k,k'} \iter{\bar{\tau}}{k}\iter{\bar{\tau}}{k'} \\+ \iter{\kappa^{\text{DP}}_{\ing{n}}}{k,k'} = 0,
\end{multline}
with
\begin{align}
    \iter{\alpha^{\text{DP}}_{\ing{n}}}{k,k'} & = 2\left(\iter{\tau_{\ing{n}}}{k} - \iter{\xi^{\text{DP}}_{\ing{n}}}{k,k'} \iter{\tau_{\ing{n}}}{k'} \right), \nonumber \\
    \iter{\chi^{\text{DP}}_{\ing{n}}}{k,k'} & = 2\left(\iter{\xi^{\text{DP}}_0}{k,k'} - \iter{\xi^{\text{DP}}_{\ing{n}}}{k,k'} \right), \nonumber \\
    \iter{\kappa^{\text{DP}}_{\ing{n}}}{k,k'} &= \left(\iter{\tau_{\ing{n}}}{k} \right)^2 + \left(\iter{\tau_{\ing{n}}}{k'}\right)^2 - 2\iter{\tau_{\ing{n}}}{k} \iter{\tau_{\ing{n}}}{k'} \iter{\xi^{\text{DP}}_{\ing{n}}}{k,k'}, \nonumber \\
    \text{and} \; \iter{\xi^{\text{DP}}_{\ing{n}}}{k,k'} &= \langle \iter{\vec{u}_{\ing{n}}}{k}, \iter{\vec{u}_{\ing{n}}}{k'} \rangle.  \nonumber
\end{align}

An appealing quality of the distance preservation condition is that it does not require any specific assumption regarding the geometry of the environment. A slightly more conservative, yet in practice still very plausible assumption is to consider all reflecting surfaces to be either horizontal (floor, ceiling, tables, etc.), or vertical (walls, windows etc.). Note that this assumption would still allow the vertical surfaces to form arbitrary angles (one such example could be an open door and its supporting wall). Advantageously, neither this nor the previous condition depend on the acoustic reflection order.

In order to exploit this \emph{horizontal/vertical (HV)} assumption, we require that the z-axis of the local coordinate system of the Ambisonic microphone is alligned with the z-axis of the global coordinate system (of the room). This generally holds true, since the microphone is most often placed on a horizontal surface (\emph{e.g.}, a table), or it is mounted on a vertical stand. If this is the case, it is evident from the Fig~\ref{figDPHV} that the magnitude of the projection of the displacement vector $\iter{\vec{r}_{\ing{n}}}{k'} - \iter{\vec{r}_{\ing{n}}}{k}$ onto the z-axis, for any $\ing{n}$, is equal to the magnitude of the same projection of the corresponding source displacement vector $\iter{\vec{r}_0}{k'} - \iter{\vec{r}_0}{k}$. Namely, for $\ing{n}=\ing{1}$, we have
\begin{equation}\label{eqProjPreservation}
    \langle \vec{u}_z, \iter{\vec{r}_{\ing{n}}}{k'} - \iter{\vec{r}_{\ing{n}}}{k} \rangle^2 = \langle \vec{u}_z, \iter{\vec{r}_0}{k'} - \iter{\vec{r}_0}{k} \rangle^2,
\end{equation}
where $\vec{u}_z = \transp{[\begin{smallmatrix} 0 & 0 & 1 \end{smallmatrix} ]}$. Developing \eqref{eqProjPreservation} from  \eqref{eq_r0} and \eqref{eq_rn}, yields
\begin{multline}\label{eqHV}
    \iter{\alpha^{\text{HV}}_{\ing{n}}}{k,k'} \iter{\bar{\tau}}{k} + \iter{\alpha^{\text{HV}}_{\ing{n}}}{k',k} \iter{\bar{\tau}}{k'} + \iter{\rho^{\text{HV}}_{\ing{n}}}{k} \left(\iter{\bar{\tau}}{k} \right)^2 + \\ \iter{\rho^{\text{HV}}_{\ing{n}}}{k'} \left(\iter{\bar{\tau}}{k'} \right)^2 + \iter{\chi^{\text{HV}}_{\ing{n}}}{k,k'} \iter{\bar{\tau}}{k}\iter{\bar{\tau}}{k'} + \iter{\kappa^{\text{HV}}_{\ing{n}}}{k,k'} = 0,
\end{multline}
with $\iter{z_{\ing{n}}}{k} = \langle \vec{u}_z, \iter{\vec{u}_{\ing{n}}}{k} \rangle$ and
\begin{align}
    \iter{\alpha^{\text{HV}}_{\ing{n}}}{k,k'} &= 2 \iter{z_{\ing{n}}}{k} \left(\iter{\tau_{\ing{n}}}{k} \iter{z_{\ing{n}}}{k} - \iter{\tau_{\ing{n}}}{k'} \iter{z_{\ing{n}}}{k'} \right), \nonumber \\
    \iter{\rho_{\ing{n}}^{\text{HV}}}{k} &= \left(\iter{z_{\ing{n}}}{k}\right)^2 - \left(\iter{z_0}{k}\right)^2, \nonumber \\
    \iter{\chi_{\ing{n}}^{\text{HV}}}{k,k'} &= 2\left(\iter{z_0}{k}\iter{z_0}{k'} - \iter{z_{\ing{n}}}{k} \iter{z_{\ing{n}}}{k'} \right), \\
    \iter{\kappa^{\text{HV}}_{\ing{n}}}{k,k'} &=  \left(\iter{\tau_{\ing{n}}}{k} \iter{z_{\ing{n}}}{k} \right)^2 +  \left(\iter{\tau_{\ing{n}}}{k'} \iter{z_{\ing{n}}}{k'} \right)^2 \nonumber \\
    & - 2\iter{\tau_{\ing{n}}}{k} \iter{\tau}{k'}_{\ing{n}} \iter{z_{\ing{n}}}{k}\iter{z_{\ing{n}}}{k'}. \nonumber
\end{align}

%Note that the distance preservation is not guaranteed under \eqref{eqProjPreservation}, hence, instead of being used independently, this condition should rather complement the one given in \eqref{eqDistPreserv}. These can be combined by \emph{e.g.}, summing the two expressions:
%\begin{multline}\label{eqDPHV}
%    \iter{\alpha_{\ing{n}}}{k,k'} \iter{\bar{\tau}}{k} + \iter{\alpha_{\ing{n}}}{k',k} \iter{\bar{\tau}}{k'} + %\iter{\rho^{\text{HV}}_{\ing{n}}}{k} \left(\iter{\bar{\tau}}{k} \right)^2 + \\ %\iter{\rho^{\text{HV}}_{\ing{n}}}{k'} \left(\iter{\bar{\tau}}{k'} \right)^2 + \iter{\chi_{\ing{n}}}{k,k'} %\iter{\bar{\tau}}{k}\iter{\bar{\tau}}{k'} + \iter{\kappa_{\ing{n}}}{k,k'} = 0,
%\end{multline}
%with $\alpha := \alpha^{\text{DP}} + \alpha^{\text{HV}}$, $\chi := \chi^{\text{DP}} + \chi^{\text{HV}}$ and $\kappa := \kappa^{\text{DP}} + \kappa^{\text{HV}}$.

The expressions \eqref{eqDP} and \eqref{eqHV} can be compactly written as %and \eqref{eqDPHV} write compactly as
\begin{equation}\label{eqBiaffine}
    \transp{ {\iter{\vect{m}_{\ing{n}}}{k,k'}} } \boldsymbol{f}(\vect{\bar{\tau}}) + \iter{\kappa_{\ing{n}}}{k,k'} = 0,
\end{equation}
where the vector $\vect{\bar{\tau}} = \transp{\left[ \begin{smallmatrix} \iter{\bar{\tau}}{1} & \iter{\bar{\tau}}{2} & \hdots & \iter{\bar{\tau}}{K}  \end{smallmatrix} \right]}$ contains the ToA estimates at all considered frames $\ing{k} \in [1, \ing{K}]$. Consequently, $\iter{\kappa_{\ing{n}}}{k,k'} \in \{ \kappa_{\ing{n}}^{\text{DP}}, \kappa_{\ing{n}}^{\text{HV}}  \}$, while the row vector $\transp{ {\iter{\vect{m}_{\ing{n}}}{k,k'}} }$ contains the remaining coefficients of a corresponding expression, along with the appropriate zero-padding. The vector-valued function $\boldsymbol{f}(\vect{\bar{\tau}})$ is defined as
\begin{equation}
    \boldsymbol{f}(\vect{\bar{\tau}}) := \left[ \begin{matrix} \vect{\bar{\tau}} \\ \vtriu{\vect{\bar{\tau}}\transp{\vect{\bar{\tau}}}} \end{matrix} \right],
\end{equation}
where $\vtriu{\vect{\bar{\tau}}\transp{\vect{\bar{\tau}}}}$ extracts the upper-triangular part of the matrix $\vect{\bar{\tau}}\transp{\vect{\bar{\tau}}}$, and concatenates its entries into a column vector.

Having assembled together (all or a subset of) frames where the same reflections have been observed, we arrive at the (usually overdetermined) nonlinear system of equations of the type \eqref{eqBiaffine}:  
\begin{equation}\label{eqBiaffineSystem}
    \mtrx{M}\boldsymbol{f}(\vect{\bar{\tau}})  + \vect{\kappa} = \left[ \begin{smallmatrix} \vdots \\ \transp{ {\iter{\vect{m}_{\ing{n}}}{k,k'}} } \\ \vdots \end{smallmatrix} \right] \boldsymbol{f}(\vect{\bar{\tau}}) + \left[ \begin{smallmatrix} \vdots \\ \iter{\kappa_{\ing{n}}}{k,k'} \\ \vdots \end{smallmatrix} \right] =  \vect{0},
\end{equation}
with $(\ing{k},\ing{k'}) \in [1,\ing{K}] \times [1,\ing{K}]$ and $\ing{n} \in [1,\ing{N}]$.

Note that the HV condition \eqref{eqProjPreservation} does not guarantee the preservation of distance  \eqref{eqDistPreserv}, hence, instead of being used independently, these should rather complement one another. This can be done by concatenating the appropriate expressions \eqref{eqBiaffine}, corresponding to each condition within the system \eqref{eqBiaffineSystem}.

%By $\mtrx{M}_{\text{DP}}$, $\vect{q}_{\text{DP}}$, and $\mtrx{M}_{\text{HV}}$, $\vect{q}_{\text{HV}}$, we denote the systems corresponding to the distance preservation property, and horizontal/vertical surfaces hypothesis, respectively. Moreover (and it would prove to be the most useful), one could promote both conditions simultaneously by stacking together the two systems into a combined one: $\mtrx{M} = \left[ \begin{smallmatrix} \mtrx{M}_{\text{DP}} \\ \mtrx{M}_{\text{HV}} \end{smallmatrix} \right]$ and $\vect{q}=\left[ \begin{smallmatrix} \vect{q}_{\text{DP}} \\ \vect{q}_{\text{HV}} \end{smallmatrix} \right]$.

Estimating the ToA vector $\vect{\bar{\tau}}$ comes down to the regression problem:
\begin{equation}\label{eqOptProblem}
    \vect{\hat{\tau}} = \argmin_{lb \leq \vect{\bar{\tau}} \leq ub} \ell \left( \diag{\vect{\psi}} \left(\mtrx{M} \boldsymbol{f}(\vect{\bar{\tau}}) + \vect{q} \right) \right) + \lambda r(\vect{\bar{\tau}}),
\end{equation}
where $0<lb < ub$ denote the lower, respectively, upper bounds for the ToA estimate. The data fidelity term $\ell(\cdot)$ is usually some type of (squared) norm, such as the sum of squares ($\ell=\ell_2^2$), or absolute values ($\ell=\ell_1$). The former leads to a smooth optimization problem, but the advantage of the $\ell_1$ norm is that it may be more robust with regards to possible errors in system parameters (the extracted DoAs and relative delays). %An alternative could be the use of \emph{structured} norms, for example, the $\ell_{1,2}$ norm, if the parameters related to certain reflections are significantly more erroneous than the others. 
Optionally, should we have some confidence indicators, the weights $\vect{\psi}$ can reflect the reliability of each condition \eqref{eqBiaffine} within the system \eqref{eqBiaffineSystem}. The regularizer $\lambda r(\vect{\bar{\tau}})$ can be applied to infuse some additional structure in the vector $\vect{\bar{\tau}}$, \emph{e.g.}, smoothness. The problem \eqref{eqOptProblem} is non-convex, whose local solution can be found by applying some nonlinear optimization method. In particular, we use the Fast Adaptive Shrinkage/Thresholding Algorithm (FASTA) \cite{FASTA:2014}, provided with the appropriate (sub)gradient of the loss function.

An attentive reader may have noticed that the necessary ingredient for the proposed method to work is that we can cluster the individual reflections across time. In other words, it is necessary to track the source's DoA and its reflections. For that purpose, we modify the tracker described in %\cite{valin2007robust,
\cite{kitic2018tramp}, such that it can process observations in the form of DoAs and the associated relative delays. A simple, but effective modification is to provide the measurements in the form of scaled vectors $\{\iter{\tau_{\ing{i}}}{k} \iter{\vec{u}_{\ing{i}}}{k}\}_{\ing{i}}$, which enables the tracker to discriminate the reflections of very similar DoAs, by adding certain ``depth'' to the observations. In practice, we run two instances of the tracking algorithm \cite{kitic2018tramp}. The first one tracks the source itself, and is, thus delegated to estimate a single target trajectory using the DoA obtained from $\vect{v}(t=0)$. The second performs multitarget tracking of the reflections using the remaining observations (directions and relative delays obtained from the remaining peaks of the GTVV sequence).

\begin{figure}
    \centering
    \includegraphics[width=\columnwidth]{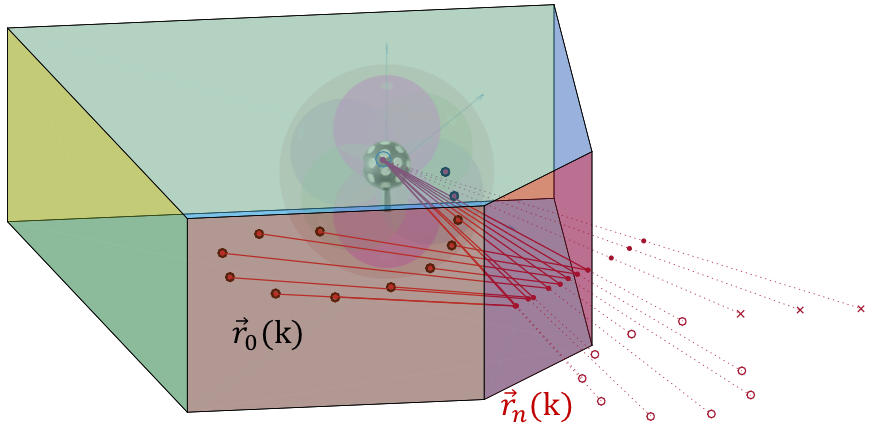}
    %\caption{An example where some of the second order reflections are not visible from the point of view of the microphone array (the big circle).}
    \caption{A source-image is "visible" ({\color{purple}o}) – or resp. not ({\color{purple}x}) –  when its virtual path to the microphone ({\color{purple}dotted line}) crosses ({\color{purple}dot ·}) the reflection plane inside – resp. outside – the panel boundaries. A reflected path ({\color{purple}cont. line}) actually hits the microphone only when the image is "visible".}
    \label{figVisibility}
\end{figure}

Finally, we remark that, when defining the two aforementioned conditions, it is possible to consider a pair of reflections $(\ing{m}, \ing{n})$ -- provided that they are both detected at two distinct frames $\ing{k}$ and $\ing{k'}$ -- instead of pairing the parameters of the source and a single reflection.  However, tracking reflections is generally less stable and accurate than tracking the source's DoA. For instance, some reflections can appear and disappear according to the current position of the source, as they can become ``invisible'' from the microphone perspective (\emph{cf}. Fig~\ref{figVisibility}). Hence, such couplings have not been considered in this work.

\section{Experiments}\label{sec:experiments}

To our best knowledge, very few methods have been proposed for evaluating the distance from the Ambisonics (or any other) microphone array to a far field sound source. Such methods usually require a prior estimate of Direct-to-Reverberation-Ratio (DRR), and can provide only a very crude guess of the true distance \cite{lu2010binaural}. In addition, in earlier works \cite{daniel2020time,kitic2021generalized} it has been shown that (G)TVV performs favorably against the beamforming-based DoA estimation \cite{kitic2018tramp}, for every HOA order $\ing{l}\in[1,4]$. Therefore, the results presented in this section do not include a baseline approach.

The proposed method is tested on the recordings from the Task~3 of the \emph{Localization And Tracking} (LOCATA) challenge \cite{evers2020locata} dataset. The 3D position of a mobile speaker (locutor) relative to the static microphone array has been provided at the rate of $120$ Hz, by an infrared optical tracking system. The room where the recording took place is modestly reverberant (T60 = 0.55s), and the signals captured by the Eigenmike\textregistered \  array were subject to measurement and ambient noise. These microphone inputs have been converted to HOA ($\ing{L}=4$) format, and downsampled to $f_s=16$~KHz. Further, we apply STFT using the Tukey window function (cosine fraction $0.5$), by dividing signals into frames of duration $0.128$~s, with $75\%$ overlap among them. The buffer allocated for the robust GTVV estimation \eqref{eqRobust} is $0.5$~s long, \emph{i.e.} $\ing{T}=16$ frames.

\begin{table}[]
    \centering
    \begin{tabular}{|c|c|c|c|}
        \hline
         DoA &      Median  &   Mean    &   St. deviation\\
         \hline 
       Azimuth      &   $7.121$ &   $8.856$ &   $6.700$   \\
       Elevation    &   $5.447$ &   $5.858$ &   $3.321$   \\
       \hline
    \end{tabular}
    
     \vspace{1em}
    
    \begin{tabular}{|c|c|c|c|c|}
        \hline 
        Hypothesis  &   \multicolumn{2}{|c|}{DP}  &   \multicolumn{2}{|c|}{DP + HV}     \\
        \hline
        Data fidelity      &   $\ell_2^2$  &   $\ell_1$    &   $\ell_2^2$  &   $\ell_1$    \\
        \hline
        Median   &   $0.601$          &   $0.340$   &   $0.732$     &   $\mathbf{0.326}$    \\
        Mean     &   $0.743$          &   $0.472$   &   $0.538$     &   $\mathbf{0.448}$    \\
        St. deviation &   $0.569$          &   $0.421$   &   $0.803$     &   $\mathbf{0.416}$    \\
        \hline
    \end{tabular}
    
    \caption{DoA estimation error, in degrees (above), and range estimation error, in meters (below).}
    \label{tab:Results}
\end{table}

% L2 DP:            0.743/0.569/0.601
% L1 DP:            0.472/0.421/0.340
% L2 DPHV:          0.803/0.538/0.732
% L1 DPHV:          0.448/0.416/0.326

\begin{figure}
    \centering
    \includegraphics[width=\columnwidth]{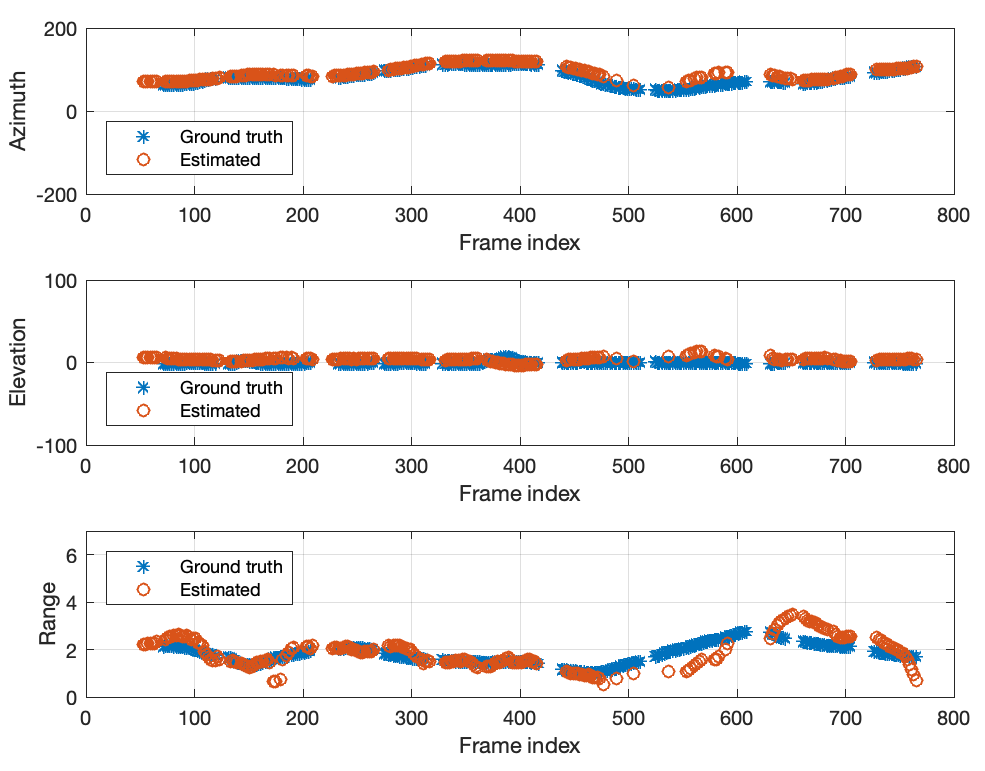}
    \caption{Estimated azimuth, elevation and range (using the $\ell_1$ norm, with DP + HV) per frame, of a recording from the LOCATA evaluation dataset.}
    \label{fig:example}
\end{figure}

The LOCATA dataset splits into validation and evaluation parts. We have used the former to tune the parameter $\lambda$ of the smoothness-promoting regularizer $\lambda \norm{\nabla\vect{\tau}}{2}^2$, with $\nabla\vect{\tau}$ being the forward difference approximation of the gradient. As a data fidelity term in the objective \eqref{eqOptProblem}, we have used either the sum of squares $\ell_2^2$, or the $\ell_1$ norm, and the estimates were constrained to the range $\bar{\tau} \in [0.5\text{m},6\text{m}]/c$. The results provided in Table~\ref{tab:Results} have been obtained by considering per-frame errors for all recordings of the evaluation dataset (the duration of each audio is approximately $25$s). The range estimation results in the lower table indicate that the proposed method provides rather accurate predictions, especially with the $\ell_1$ norm in the loss function. The performance further improves when both DP and HV hypothesis are simultaneously exploited. A visual example of the prediction is given in Fig.~\ref{fig:example}.

\section{Conclusion}\label{sec:conclusion}

We have addressed the challenging problem of 3D localization of a mobile, far field sound source. The proposed method is capable of per-frame estimation of azimuth, elevation and range, by exploiting acoustic reflections whose parameters are inferred directly from the GTVV representation of an Ambisonic recording. The method is based on geometrical principles, and very general assumptions on the acoustic environment. Despite its simplicity, the method is surprisingly efficient, as demonstrated on the real audio data from the LOCATA challenge. The future work will focus on improving its robustness, \emph{e.g.} in the multiple source scenario. 

\bibliographystyle{IEEEtran} % We choose the "plain" reference style
\bibliography{EUSIPCO22} % Entries are in the refs.bib file

\end{document}